# PUMiner: Mining Security Posts from Developer Question and Answer Websites with PU Learning

Triet Huynh Minh Le[1], David Hin[1,2] Roland Croft[1], and M. Ali Babar[1,2]
[1]School of Computer Science, The University of Adelaide, Adelaide, Australia
[2]Cyber Security Cooperative Research Centre
Email: triet.h.le@adelaide.edu.au, david.hin@student.adelaide.edu.au,
roland.croft@student.adelaide.edu.au, ali.babar@adelaide.edu.au

*Abstract*— Security is an increasing concern in software development. Developer Question and Answer (Q&A) websites provide a large amount of security discussion. Existing studies have used human-defined rules to mine security discussions, but these works still miss many posts, which may lead to an incomplete analysis of the security practices reported on Q&A websites. Traditional supervised Machine Learning methods can automate the mining process; however, the required negative (non-security) class is too expensive to obtain. We propose a novel learning framework, PUMiner, to automatically mine security posts from Q&A websites. PUMiner builds a context-aware embedding model to extract features of the posts, and then develops a two-stage PU model to identify security content using the labelled Positive and Unlabelled posts. We evaluate PUMiner on more than 17.2 million posts on Stack Overflow and 52,611 posts on Security StackExchange. We show that PUMiner is effective with the validation performance of at least 0.85 across all model configurations. Moreover, Matthews Correlation Coefficient (MCC) of PUMiner is 0.906, 0.534 and 0.084 points higher than one-class SVM, positive-similarity filtering, and one-stage PU models on unseen testing posts, respectively. PUMiner also performs well with an MCC of 0.745 for scenarios where string matching totally fails. Even when the ratio of the labelled positive posts to the unlabelled ones is only 1:100, PUMiner still achieves a strong MCC of 0.65, which is 160% better than fully-supervised learning. Using PUMiner, we provide the largest and up-to-date security content on Q&A websites for practitioners and researchers.

## I. INTRODUCTION

Security incidents and their damaging effects are increasing at an unprecedented rate [1]. There are several platforms reporting security vulnerabilities and attacks such as Common Vulnerabilities and Exposures (CVE) [2], National Vulnerability Database (NVD) [3], Common Weakness Enumeration (CWE) [4], Common Attack Pattern Enumeration and Classification (CAPEC) [5], and Open Web Application Security Project (OWASP) [6]. These platforms provide definitions and examples of known security attacks and countermeasures. However, they do not support on-the-fly discussion about the up-to-date usage and implementation of such security in real-life scenarios. For example, CWE suggests using the *mysql_real_escape_string()* function to mitigate SQL injection vulnerability in PHP [7], but this function was deprecated and one should use PDO or MySQLi instead for better protection [8]. Conversely, developer Question and Answer (Q&A) websites provide an abundance of such discussions [9]. Until 2019, Stack Overflow [10], the largest developer Q&A site, has had around seventeen million posts and more than ten million users. This is an enormous data source for mining patterns in secure software development.

Several studies [11-14] have proposed different heuristics to retrieve security content from open sources. Such predefined rules require considerable domain expertise and human effort. More importantly, the rules are not exhaustive since security concepts keep evolving over time [15], see section II.B for more challenges with the existing heuristics. There is a need for an automatic and accurate way of retrieving security content. It is observed that recent unsupervised word/document embeddings [16-19] in Natural Language Processing (NLP) have achieved state-of-the-art results for various textual information retrieval tasks. Such NLP methods can then be combined with a Machine Learning (ML) classifier to automatically identify security-related posts on developer Q&A websites without human-defined rules.

Building a traditional binary classification model requires a large amount of labelled data containing both positive (security) and negative (non-security) classes [20, 21]. However, it is very challenging to obtain such high-quality and labelled datasets. Even with our sophisticated heuristics using the knowledge from multiple security sources (e.g., NVD, CVE, CWE, CAPEC, and OWASP), we are still unsure of the labels of more than 98% of the posts, leaving them unlabelled. Human inspection also does not scale to such a large extent. Fortunately, there are still related yet much smaller sources that are likely to be related to the domain of interest for learning the patterns of the positive class. In the context of security, Security StackExchange [22] contains mostly security-related discussion, and this site also follows the same format as Stack Overflow. Such observations have motivated us to formulate and address a research problem "**How to retrieve security-related content from developer Q&A websites using only a small quantity of labelled positive posts and a large number of unlabelled posts?**"

We present **PU**Miner, a novel learning framework to mine the posts belonging to a topic of interest from developer Q&A websites using only **P**ositive and **U**nlabelled posts. Specifically, we represent the posts with a context-aware feature model and then build a two-stage PU learning model to discern security-related posts on developer Q&A websites. Such security posts help researchers and practitioners study security practices/guidelines in a software development lifecycle as well as improve the discovery and mitigation of software vulnerabilities. We have conducted large-scale experiments on more than 17.2 million Stack Overflow posts and 52,611 Security StackExchange posts. Our findings are:

- PUMiner is effective for retrieving security posts as all of its models achieved at least 0.85 F1-Score and G-mean on a PU validation set.
- PUMiner had a testing Matthews Correlation Coefficient (MCC) value that was 0.906, 0.534 and 0.084 points better than one-class SVM [23], positive-similarity filtering [24] and one-stage PU mod-els. PUMiner also had the best MCC of 0.745 for cases where keyword matching completely failed.

- PUMiner consistently outperformed a fully supervised approach treating all unlabelled data as negative class. Even when the unlabelled posts were 100 times more than the labelled positive ones, PUMiner still achieved a strong testing MCC of 0.65.

In brief, our noteworthy contributions are:

1. We are the first to formulate the retrieval of security-related posts as a PU learning problem.
2. We propose a novel framework, PUMiner, to retrieve security posts using unlabelled data.
3. We systematically evaluate the performance of PUMiner using more than 17.3 million posts on both Stack Overflow and Security StackExchange.
4. We provide the largest security-related datasets extracted from Stack Overflow and code at [25].

**Outline**. Section II introduces (security) post discussion on Q&A websites and motivates PU learning. Section III describes the proposed PUMiner framework. Section IV presents our study design/setup. Section V reports the experimental results. Section VI discusses the findings and threats to validity. Section VII mentions the related work. Section VIII summarizes this work and suggests future directions.

## II. BACKGROUND AND MOTIVATIONS

### A. Security discussion on question and answer websites

A discussion thread on Stack Overflow developer Question and Answer (Q&A) website is called a post [26]. Fig. 1 illustrates the format of a post. A post contains the following main components: a unique id, owner, title, question, tag(s) and answer(s). Among the answers, one would be selected as the best one by the post questioner. A post also has additional information such as comments, the dates it when was created, edited, last-active and closed, along with the score (upvotes minus downvotes), the number of views and favorites. A complete schema of a post on Q&A websites can be found at [27]. Security StackExchange, a security Q&A website, also follows a similar format as Stack Overflow.

To identify security-related posts, our work would consider the title, tags, question and answer bodies following the practices of the previous studies [11, 12, 26]. The security relevance of a post is then decided based on the explicit or implicit reference or discussion about cybersecurity and/or any related subtopics mentioned in the title, question and answers of a post. A post on how to use/develop security tools/functions/frameworks (e.g., login/authentication) is still related to security. However, discussing a non-security task (e.g., integrating third-party APIs[1]) that may indirectly involve security requirements (e.g., user's authentication/permission) is not security-relevant.

Security has become an increasingly important topic in Software Engineering [28]. However, the number of security posts identified on Stack Overflow in the literature is still very limited. From our manual inspection of 385 random posts (i.e., 95% confidence level and 5% error [29]), we found sixteen (around 4%) posts were security-related.

---

[1] https://stackoverflow.com/questions/35275411/trying-to-connect-to-facebook-through-the-api

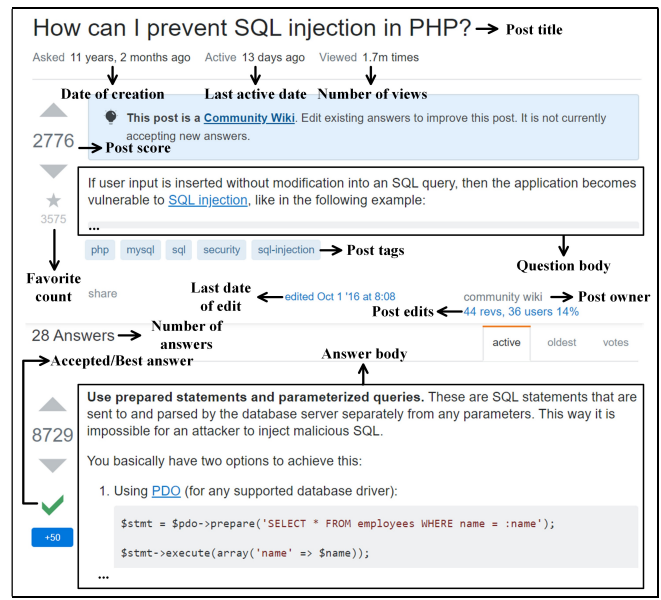

Fig. 1. Format of a (security) post (id 60174) on Stack Overflow.

However, the largest set [12] available in the literature just contains around thirty thousand posts (only about 0.1% of 17.2 million Stack Overflow posts). This set of posts was created with a keyword-matching approach using only the tags (i.e., *encryption*, *cryptography*, *passwords*, *security*, *sql-injection*, *web-security*, *xss*) of a post. There are many cases in which a security post does not have any of these seven tags. One such example is the post with question id 3226374 on Stack Overflow[2], discussing a cross-site scripting issue in PHP using only the tag *php*. More limitations of keyword matching are discussed in the next section.

### B. Keyword matching and its limitations

The most straightforward way to determine the type of a post is by matching it with predefined keywords in the domain of interest (i.e., security). However, there are two major drawbacks of such a keyword-matching approach. Firstly, it requires considerable domain expertise and a tedious trial-and-error process to select appropriate keywords. For example, to our best knowledge, the most extensive list has security 127 keywords proposed by Pletea et al. [14]. However, based on our expertise, we have found out that such list is still missing many important security keywords. With further investigation, we found out that missing the keyword "*ssh*" (a cryptographic network protocol) alone would result in 32,216 potential posts being overlooked.

Another important pitfall of keyword matching is that it is hard to balance between Precision (accurate information retrieval) and Recall (complete information retrieval). With fewer keywords, it is likely to miss many posts of interest (i.e., low recall) as demonstrated above. On the contrary, even with more relevant keywords, it does not guarantee a better matching accuracy. In fact, more keywords may result in a higher false-positive rate due to the multiple meanings of a word. For example, the word "*exploit*" in the previous security list [14] also means "*take advantage of*" some techniques/tools in general, which is not necessarily related to a security attack.

---

[2] https://stackoverflow.com/questions/3226374/cross-site-scripting-phishing-through-frames

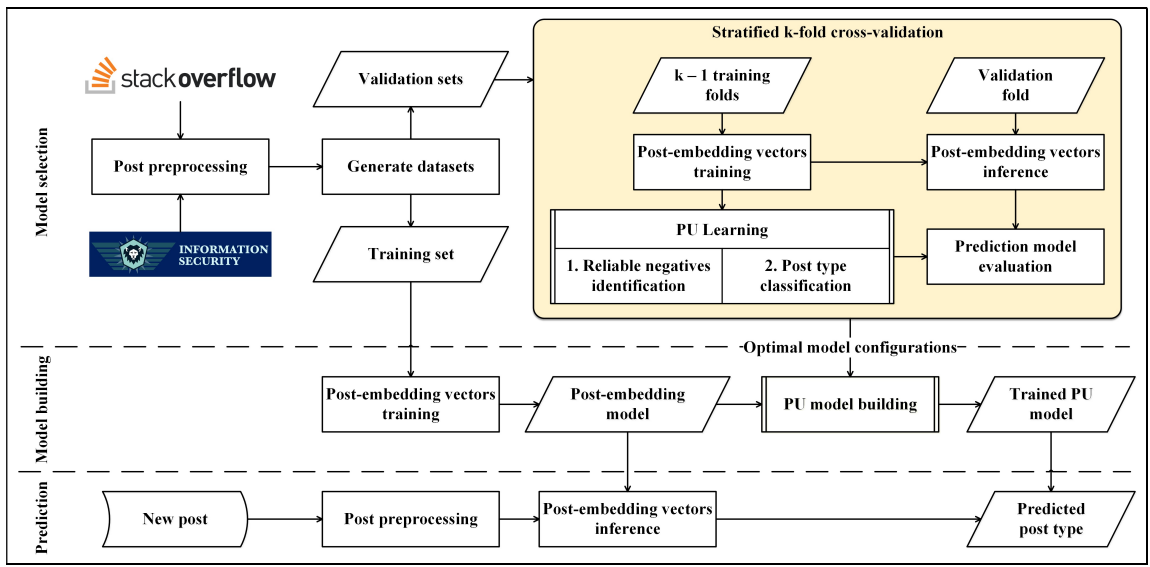

Fig. 2. Workflow of our proposed PUMiner framework for mining security-related posts from developer Q&A websites.

Another example is the word "*signed*", which is also a matching subword of non-security words like "*assigned*". Therefore, a more reliable and generalized technique is required for addressing such drawbacks.

### C. The need for PU learning for retrieval of security posts

Machine Learning (ML)-based binary classifiers can automatically identify important keywords to determine a type of interest. To build a binary classifier, we need to have both positive (i.e., security-related) and negative (i.e., non-security related) posts. Unfortunately, there has not been such a large and reliable dataset containing both positive and negative classes in the literature. We have only identified a small number of positive posts with high confidence from Stack Overflow and Security StackExchange. Retrieving non-security posts is also challenging since it should not contain any security context, which requires significant human effort to define and verify. Hence, there are many posts still left unlabelled (i.e., containing a mixture of positive and negative samples). This would violate the prerequisite condition of developing a traditional binary classifier since the negative class is unknown [21, 30]. Nonetheless, this is a suitable case for **PU** learning to utilize both available **P**ositive and **U**nlabelled samples [31]. Such observations have motivated us to formulate the identification of security posts on Q&A websites as a PU binary learning problem.

Input representation is another major concern when building an ML model. In NLP, a textual post on a Q&A website can be represented using Bag-of-Words (BoW), n-grams, or tf-idf. However, these traditional text representation techniques only support the exact matching of each word without considering its semantic meaning [32, 33]. For instance, "*penetration-testing*" or "*pen-test*" is usually used for vulnerability assessment (i.e., security-related context), but BoW, n-grams or tf-idf would treat "*pen-test*" and "*security*" as two totally different words. To address such drawback, word/document embeddings [16, 19] are utilized to incorporate the context (neighboring words) of a considering word. Therefore, our work would use context-aware word/document embeddings to represent the content of a discussion post to determine its security relevance.

## III. THE PUMINER FRAMEWORK

### A. Approach Overview

We propose **PU**Miner to automatically perform large-scale mining of security content from developer Q&A websites using **P**ostive and **U**nlabelled posts. The overall workflow of our proposed PUMiner is described in Fig. 2. There are three processes in PUMiner framework: (*i*) model selection, (*ii*) model building, and (*iii*) prediction. The model selection and building processes develop the optimal PU learning models, and the prediction process uses such models to predict the security relevance for a new post. It is noted that the workflow of PUMiner can be customized to retrieve any topic of interest from developer Q&A websites.

**Model selection**. There are three steps: (*i*) preprocess posts from Q&A websites, (*ii*) generate datasets for validation and training, and (*iii*) perform stratified k-fold cross-validation to select the best hyperparameters for our PU models. Step (*i*) accepts and preprocesses raw posts from Stack Overflow (the largest developer Q&A website) and Security StackExchange (security-focused Q&A site). Preprocessing techniques used for the title, question body, tags and answer of a post are covered in section III.B. Step (*ii*) takes the preprocessed posts to create a reliable positive (security) dataset for PU learning. Although PU learning does not require the negative class, it still needs to learn the patterns from the positive class. We propose different heuristics in section III.C for building a positive dataset using both the tags and content of post title, question and answers. Such heuristics have been defined based on our expertise and references from CVE, NVD, CWE, CAPEC and OWASP. It should be noted that these heuristics are insufficient for retrieving security posts at full scale as described in section II.B. Besides the labelled positive set, step (*ii*) also obtains an unlabelled set to build a complete dataset for validating/training PU models. The validation set is separated into *k* folds. Each fold contains both labelled as security and unlabelled posts.

The folds enter the last step (*iii*) to perform stratified k-fold cross-validation. Stratification ensures that the ratio of each input source is kept throughout the cross-validation step, avoiding different data distribution of the folds.

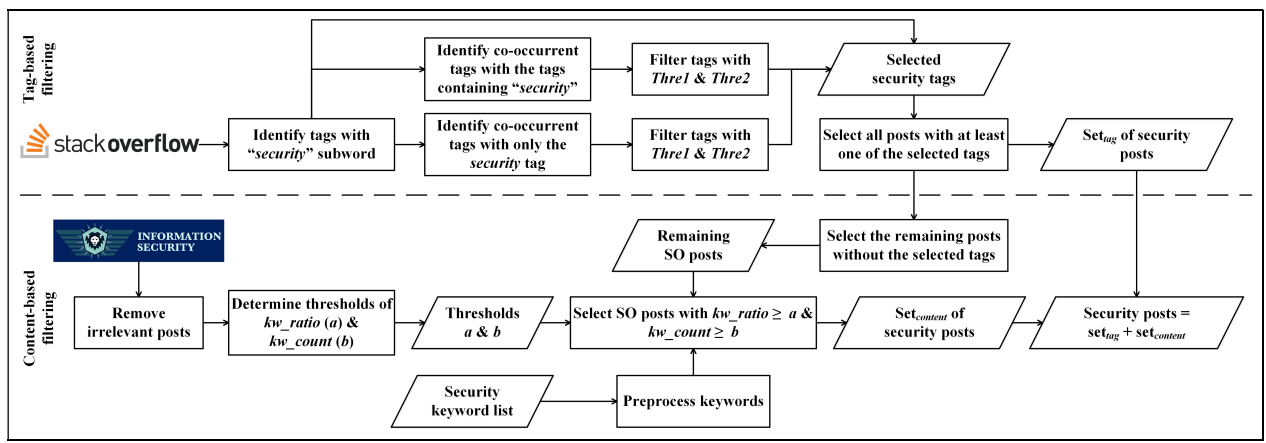

Fig. 3. Workflow for building a security dataset using tag-based and content-based heuristics.

In each iteration, $k - 1$ folds are used for training a model, while the remaining one is used for validating such model. In the training step of an iteration, textual posts and their tags are first fed into a doc2vec [19] embedding model to obtain fixed-length context-aware feature vectors. The embedding model is also stored temporarily for inferring the vectors of the posts in both the training and validation folds. The trained feature vectors of the training folds then enter a two-stage PU learning model (see section III.D). The first step of PU learning identifies the reliable negative (non-security) posts in the training folds to approximately transform this PU learning problem into a fully-supervised learning one. After that, both positive (security) and reliable negative (non-security) posts in the training folds are used to train a traditional binary classifier to predict the type of each post. Such trained model is then evaluated on the validation fold. This whole training-validation process is repeated $k$ times. The validation performance of a model is the average value of $k$ runs. The model configurations with the highest value of performance metric would be selected as the optimal ones for the next model building process.

**Model building.** This process has two steps: (*i*) train a feature model to obtain feature vectors using doc2vec embedding model, and (*ii*) train a PU learning model to predict the security relevance of a post with the optimal configurations from the previous process. In step (*i*), doc2vec embedding is trained on the whole dataset generated in step (*ii*) of the model selection process to obtain the context-aware feature vectors of each word in the posts and the posts themselves. This feature model is also saved to disk for looking up the vectors of future posts. Using the obtained features, step (*ii*) trains a two-stage PU learning model with the optimal configurations of the model selection process. Lastly, such prediction model is then saved to disk for future inference.

**Prediction.** This process first reuses the same preprocessing module to clean a new post. The cleaned content is then converted into context-aware vectors by the feature model developed in the model building process. The trained PU learning model would use such features to determine whether the current post is related to security. We do not use tags for prediction since they may not capture the true topic of a post (e.g., the post 3226374 in section II.A).

### B. Preprocessing of discussion posts

Discussion posts on Q&A websites contain too much noise for a learning model to be applied directly. The main problem with raw posts is the large vocabulary of the post content, which can make a learning model overfit [34]. Following the practices of the previous works [12, 13, 26], we first remove code snippets and outputs (within <pre><code> and </pre></code> tags) and HTML tags (e.g., <p> and </p>). Then, we remove stop words as well as punctuations. We leave security and software engineering keywords intact, e.g., "*sql-injection*", "*private-key*", or "*x.509*". Moreover, we convert text to lowercase and perform Porter stemming [35] to remove duplicate forms of a word (e.g., "*attack*" vs. "*attacks*" vs. "*Attacker*"). We apply the same process for the title, question, tags, and answers of a post.

### C. Heuristics for building a security-related dataset

Heuristics may miss many security posts (see section II.B), but they are still useful for gathering a decent positive dataset for PU learning. We propose a unified workflow in Fig. 3 for building a security-related dataset using both **tag-based** and **content-based filtering**. Regarding the **tag-based filtering**, inspired by Yang et al. [12], we select the tags whose frequency in the security context ($|tag_i \cap security|/|tag_i|$, $|.|$ is the tag count) and popularity ($|tag_i \cap security|/|security|$) are larger than *Thre1* and *Thre2* thresholds, respectively. However, [12] only used the *security* tag to define the security context, while many other tags with "*security*" as subword (e.g., *spring-security*, *firebase-security* or *android-security*) are also related to security on Stack Overflow. Therefore, we consider all tags with the subword "*security*" along with the *security* tag to select co-occurrent tags. We have adopted *Thre1* = 0.1 and *Thre2* = 0.01 as in [12] and obtained five tags: *cryptography*, *csrf*, *passwords*, *sql-injection* and *xss*. We then retrieve all the posts with at least one of such tags to create the security set$_{tag}$. However, there is no standard procedure to assign tags [26] and there are still security posts with no security tag (e.g., post 3226374 in section II.A), requiring the examination of the content of other parts besides tags as well.

Moving on to the **content-based filtering** for the remaining posts, we present the most up-to-date list of 234 security keywords (see Appendix), which is two times larger than [14]. Such keywords are inherited from the existing list and gathered from the entities/frequent words in CVE, NVD, CWE, CAPEC and OWASP. We consider different forms (i.e., American/British English and with/without hyphen/space/stemming) of a keyword for matching to cope with various (mis-)spellings. For example, variants of "*improper synchronization*" include: improper/improp(-)synchronisation/synchronization/synchron/synchronis. Like [14], to reduce false positives, we perform exact matching for short (three-character) keywords (e.g., *md5*) and subword matching otherwise. For each post, we then extract the ratio of security words (*kw_ratio*), and how many security keywords appear in total (*kw_count*). *kw_ratio* ensures that a security word is not an outlier in the discussion. For instance, the post 477816 on Stack Overflow has the word "*security*" in the question, but the discussion is mainly about JSON MIME type. A keyword may also have several meanings, some of which are unrelated to security. For example, "*hash*" appears frequently in a post, but it refers to a data structure instead of cryptography. *kw_count* addresses such issue by increasing the coverage of security discussion in a post. We have obtained the thresholds $a = 0.053$ and $b = 8$ for *kw_ratio* and *kw_count*, respectively, from unclosed Security StackExchange posts. We have selected such posts to ensure the security relevance. Using $a$ and $b$, the content-based filtering selects Stack Overflow posts whose *kw_ratio* $\geq a$ and *kw_count* $\geq b$ to obtain set$_{content}$ of security posts.

We combine set$_{tag}$ and set$_{content}$ to form our security dataset from Stack Overflow. We have randomly sampled 385 posts (statistically significant size [29]) from each set$_{tag}$ and set$_{content}$ for two authors to independently examine the security relevance. The disagreement proportions were only 1.6% and 0.3% for set$_{tag}$ and set$_{content}$, respectively. When there was a conflict, the third author would be involved in the decision-making process. For set$_{tag}$, we encountered only two unclear security cases (< 1%), i.e., the posts 46178245 and 10727000 did not explicitly mention the security purposes for sanitizing input. For set$_{content}$, all 385 posts checked were security-related. To extend PUMiner to other domains, relevant anchor tags of tag-based filtering and keywords of content-based filtering need to be updated accordingly.

### D. Context-aware two-stage PU learning model for mining security posts

Traditional feature extraction methods like BoW, n-grams, tf-idf only treat each word as a separate unit, without considering their semantic relationship [36, 37]. Word embeddings such as word2vec [16] have been proposed to incorporate the context of a word by maximizing the following likelihood: $\prod_{t}^{N} \prod_{-ws \leq i \leq ws, i \neq t} P(w_{t+i} | w_t; \theta)$. More specifically, word2vec optimizes its parameters ($\theta$) to achieve the maximum probability of the context words appearing within a window size (*ws*) of the current word ($w_t$). Doc2vec [19] extends word2vec to jointly learn the representation of a document (paragraph vector) with its constituent words. Our work adopts the distributed memory architecture to train doc2vec model as suggested in [19]. Regarding the label of a document/post, a post may contain multiple tags to describe complex concepts, e.g., *php* and *security* tags represent security issues (sql-injection) in PHP. Therefore, besides single tags, we also combine all tags to handle the topic mixture. For example, a post with *php* and *security* tags would have *php_security* alongside *php*, *security* and *post id* as labels. We also sort labels alphabetically to avoid duplicates (e.g., *php_security* and *security_php* are the same).

**Algorithm 1.** Context-aware two-stage PU learning model building.

**Input:** List of posts: $P_{in}$
Labels of posts: $labels \in \{positive, unlabelled\}$
Size of embeddings and Window size: $sz, ws$
Classifier and its model configurations: $C, config$
**Output:** The trained feature and PU models: feature_model, model$_{PU}$

1  $word\_list, tag\_list \leftarrow$ empty list, empty list
2  **foreach** $p_i \in P_{in}$ **do**
3   $words, tags \leftarrow$ tokenize $(p_i)$, extract_tags $(p_i)$
4   $word\_list, tag\_list \leftarrow word\_list + \{words\}, tag\_list + \{tags\}$
5  **end foreach**
6  feature_model $\leftarrow$ train_doc2vec($word\_list, tag\_list, sz, ws$)
7  $\mathbf{X}_{in} \leftarrow$ obtain_feature(feature_model, $word\_list$)
8  $P \leftarrow \{\mathbf{x}_p | \mathbf{x}_p \in \mathbf{X}_{in} \wedge label(p) = positive\}$
9  $U \leftarrow \{\mathbf{x}_p | \mathbf{x}_p \in \mathbf{X}_{in} \wedge label(p) = unlabelled\}$
10 $centroid_P, centroid_U \leftarrow \dfrac{\sum_{p \in P} \mathbf{x}_p}{|P|}, \dfrac{\sum_{p \in U} \mathbf{x}_p}{|U|}$
11 $RN \leftarrow$ empty list
12 **foreach** $\mathbf{x}_i \in \mathbf{X}_{in}$ **do**     // Stage-1 PU: Identify reliable negatives
13  **if** $d(\mathbf{x}_i, centroid_U) < \alpha * d(\mathbf{x}_i, centroid_P)$ **do**
14   $RN \leftarrow RN + \{\mathbf{x}_i\}$
15  **end if**
16 **end foreach**
17 model$_{PU} \leftarrow$ train_classifier($C, P, RN, config$) // Stage-2 PU
18 **return** feature_model, model$_{PU}$

Next, we present the context-aware two-stage PU learning model to predict the security relevance of a post – the core of our PUMiner framework. Algorithm 1 describes the end-to-end PU model building. This algorithm requires a list of discussion posts with their respective labels (positive or unlabelled), along with the configurations of doc2vec (size of embeddings and window size) and classification models (model hyperparameters). The details are given hereafter.

**Lines 1-9: Learning context-aware feature vectors with doc2vec.** Lines 1-5 tokenize text and extract tags of each post to prepare the data for training doc2vec models. Line 6 trains a doc2vec embedding model with stochastic gradient descent to learn the context of words and posts. Line 7 then obtains the feature vector of each post using the trained doc2vec model. Lines 8 and 9 extract the embedding vectors of both positive and unlabelled posts, respectively.

**Lines 10-16: Stage one of PU learning model.** Inspired by the existing studies of PU learning in other domains [38-41], we assume that the context-aware doc2vec embeddings can make posts of the same class stay in close proximity in the embedding space. Such assumption has been shown to hold in section V.A. Stage-one PU learning first identifies (reliable/pure) negative (non-security) posts in the unlabelled set that are as different as possible from the positive set. A traditional binary classifier would work poorly in this case since the negative class is not pure [40]. This has been confirmed in section V.C. In line 10 of Algorithm 1, we propose to approximately locate unknown negative posts using the centroid (average) vectors of the known positive (**centroid**$_P$) and unlabelled (**centroid**$_U$) sets, respectively. Since the number of non-security posts is dominant (i.e., up to 96% on Stack Overflow, see section II.A), **centroid**$_U$ would represent the negative class more than the positive one. Lines 11-16 compute and compare the cosine distances [16, 19, 40] (see Eq. (1)) from each post to the two centroids. If the current post is closer to **centroid**$_U$ (i.e., more towards the negative class), it would be selected as a reliable negative.

$$\text{cosine\_distance} = d(i, j) = 1 - \frac{\mathbf{p}_i \cdot \mathbf{p}_j}{\|\mathbf{p}_i\| \times \|\mathbf{p}_j\|} \quad (1)$$

where $\mathbf{p}_i$ and $\mathbf{p}_j$ are the embedding vectors of the posts $i^{th}$ and $j^{th}$, respectively. The range of cosine_distance is [0, 2]. We also propose a scaling factor ($\alpha$) to increase the flexibility of our centroid-based approach, which would be jointly optimized with other hyperparameters of binary classifiers in the second stage. Besides having only one hyperparameter for tuning, this centroid-based approach can incrementally learn a new post or a set of new posts very fast with $O(1)$ complexity as given in Eq. (2).

$$\mathbf{centroid}_{new} = \frac{\mathbf{centroid}_{old} * N + \mathbf{x}_{new}}{N + \text{size}(\mathbf{x}_{new})} \quad (2)$$

where **centroid**$_{old}$, **centroid**$_{new}$ are the centroid vectors before and after learning new post(s) ($\mathbf{x}_{new}$), while $N$ is the original number of posts in the positive or unlabelled set.

**Lines 17-18: Stage two of PU learning model.** Using positive and (reliable) negative posts from the first stage, the second stage of PU learning (line 17) trains a binary classifier with its hyperparameters. In the model building process, PU model is trained with the optimal classifier and its configurations obtained from stratified k-fold cross-validation (see Fig. 2). Finally, line 18 saves the trained feature and PU models to disk for future inference.

## IV. EXPERIMENTAL DESIGN AND SETUP

### A. Dataset

We gathered 17,278,709 posts on Stack Overflow (SO) as of July 2019 from BigQuery [42], and 52,611 posts on Security StackExchange (SSE) as of October 2019 from StackExchange Data Explorer [43]. Table 1 summarizes the number of tags and the word length of each part (i.e., title, question and answer) of posts from the two above Q&A sites. SSE and SO do not only have a similar format, but their data distributions also nearly resemble, except answers. We obtained 102,046 and 27,424 security posts for set$_{tag}$ and set$_{content}$ (see Fig. 3), respectively. Set$_{tag}$ was three times larger than the largest (30,054) previous set [12] and set$_{content}$ also had nearly the same size as [12], proving that our new heuristics and keywords are useful. We then randomly sampled 181,696 posts from the remaining 17,149,239 unlabelled ones to build the *development set* along with the security posts from set$_{tag}$, set$_{content}$ and SSE (52,226 posts). 385 random SSE posts with $kw\_count = 0$ were put aside for a separate testing set, which will be explained later in this section. The *development set* had a ratio of one to one between security and unlabelled posts. 90% of the *development set* was then randomly selected for model selection and model building, and the other 10% (Test$_{dev}$) was for model prediction on unseen posts (see Fig. 2). Stratified sampling ensures that the proportion of each source would be kept.

We used the whole *development set* to test our models on two more sets (Test$_{norm}$ and Test$_{hard}$), each containing 385 (95% confidence level and 5% error [29]) manually-selected posts for each of the classes (security and non-security).

TABLE 1. STATISTICS OF POSTS ON STACK OVERFLOW AND SECURITY STACKEXCHANGE.

| Source (no. of posts) | Statistic | Tag | Title | Question | Answer |
|---|---|---|---|---|---|
| Stack Overflow (17,278,709) | Average | 3.0 | 8.6 | 159.6 | 140.8 |
| | Median | 3.0 | 8.0 | 119.0 | 87.0 |
| | Min | 1.0 | 1.0 | 1.0 | 0.0 |
| | Max | 6.0 | 45.0 | 11,025.0 | 45,004.0 |
| Security StackExchange (52,611) | Average | 2.6 | 9.2 | 140.0 | 341.4 |
| | Median | 3.0 | 9.0 | 106.0 | 225.0 |
| | Min | 1.0 | 1.0 | 1.0 | 0.0 |
| | Max | 5.0 | 30.0 | 3,714.0 | 9,069.0 |

The manual selection was conducted by one author and then carefully verified by two other authors. If no agreement could be reached, another post would be selected for examination until the size of each testing set was met. Specifically, Test$_{norm}$ had 385 security posts and 385 non-security ones manually selected from the remaining unlabelled posts that were not in the *development set*. Test$_{norm}$ was necessary since it had manually-checked label for each post, which facilitated an evaluation using both positive and negative classes. Test$_{hard}$ contained 385 positive (security-related) posts with $kw\_count = 0$ on SSE (see section III.C) and 385 negative (non-security) posts with $kw\_count > 0$ on SO that were not in the *development set*, set$_{tag}$, set$_{content}$ or Test$_{norm}$. Test$_{hard}$ had extreme cases where our tag-based and content-based keyword heuristics in section III.C totally failed. One such borderline security post is a concern about an app tracking user's location on smartphone, which affects user's privacy[3]. We have released all datasets of this work at [25].

### B. Implementation of the PUMiner framework

We have implemented our heuristics and PU learning models using *scikit-learn* [44], *nltk* [45] and *Gensim* [46] libraries in Python. The hyperparameters we considered for optimizing each model are given in Table 2. For doc2vec, we just used the default configurations since they do not make much of a difference as found in the previous studies [15, 36]. In addition, the six classifiers were selected based on the common practices in the literature [15, 47, 48] and in real data science competitions (e.g., Kaggle [49]). The first three (LR, SVM and KNN) are single models, while the other three (RF, XGB and LGBM) are ensemble ones. There were 260 model configurations for tuning in total. To select the optimal configurations for each model, we performed stratified 10-fold cross-validation. Due to the large scale of the dataset, we ran each fold as a computing job on a supercomputing cluster with at least ten cores and 400 MB of RAM. After that, we aggregated the results from multiple jobs to compute the performance of each cross-validation run. Such performance metrics are given in the next section.

### C. Evaluation metrics

Retrieving security posts is a binary classification problem, which can be evaluated using: True Positive (*TP*), False Positive (*FP*), True Negative (*TN*), and False Negative (*FN*) [20]. There are also unlabelled posts in this work; thus, we present evaluation metrics for two different scenarios in Table 3: the *development set* (only Positive and Unlabelled (PU) posts are available), and the testing sets (Positive and Negative (PN) ground truths are known).

---
[3] https://security.stackexchange.com/questions/190107/location-tracking

TABLE 2. HYPERPARAMETER TUNING FOR PUMINER.

| Modeling step | Model | Hyperparameters |
|---|---|---|
| Context-aware feature embedding | Doc2vec [19] | Embedding size: 300 Window size: 5 |
| Identification of reliable negatives | Centroid-based approach (section III.D) | Alpha ($\alpha$): 0.8, 0.9, 1.0, 1.1, 1.2 |
| Prediction of the security relevance of a post | Logistic Regression (LR) [50] | Regularization coefficient: 0.01, 0.1, 1, 10, 100 |
| | Support Vector Machine (SVM) [51] | |
| | K-Nearest Neighbors (KNN) [52] | Number of neighbors: 11, 31, 51 Distance weight: uniform, distance-based Distance norm: 1, 2 |
| | Random Forest (RF) [53] | Number of estimators: 100, 300, 500 Max depth: unlimited Maximum number of leaf nodes: 100, 200, 300, unlimited (RF) |
| | Extreme Gradient Boosting (XGB) [54] | |
| | Light Gradient Boosting Machine (LGBM) [55] | |

We estimated $r = 0.025$ in Precision$_{PU-UB}$ by manually examining 385 posts in the unlabelled set. Eq. (3) shows the relationship between G-mean values of fully-supervised learning (G-mean$_{PN}$) and PU learning (G-mean$_{PU}$).

$$\text{G-mean}_{PN} = \sqrt{P \times R} \propto \frac{P \times R}{P(y=1)} = \frac{P \times R^2}{R \times P(y=1)}$$

$$= \frac{P(y=1|\hat{y}=1) \times R^2}{P(\hat{y}=1|y=1) \times P(y=1)} \quad \text{(Definition of P and R)} \quad (3)$$

$$= \frac{R^2}{P(\hat{y}=1)} = \frac{|\hat{y}| \times R^2}{|\hat{y}=1|} = \text{G-mean}_{PU} \quad \text{(Bayes' rules, } |.|:\text{size)}$$

where P and R are Precision and Recall, respectively.
G-mean is directly proportional to Precision and Recall, similar to F1-Score. Also, $P(y=1)$ is a constant for the same dataset; hence, the model ranking using G-mean$_{PN}$ is the same as G-mean$_{PU}$. Therefore, we used G-mean$_{PU}$ as the determinant metric to select the optimal PU learning model. G-mean$_{PU}$ is the best metric for PU learning, but its value can be arbitrarily larger than one. Thus, we needed other metrics in Table 3 to provide a more human-interpretable yet less accurate evaluation of PUMiner. For evaluating on the ground-truth datasets, we also used MCC since it considers the results of both the positive and negative classes.

## V. RESEARCH QUESTIONS AND RESULTS

### A. RQ1: *Is PUMiner effective for mining security posts from developer Q&A websites?*

**Motivation.** Our work formulates the retrieval of security posts on Q&A websites as a PU learning problem. RQ1 evaluates the first attempt to address this new problem with PUMiner. Our framework allows researchers and practitioners to gather security information without the negative class.
**Approach.** We first visualize the doc2vec feature vectors using Principal Component Analysis (PCA) [21]. We then perform stratified 10-fold cross-validation and report the results of different configurations of PUMiner (see Table 2) with respect to the first six evaluation metrics in Table 3.
**Results.** We first showed that doc2vec context-aware embeddings could capture the transition from the non-security to security regions (i.e., yellow to purple based on the security keyword count ($kw\_count$)) in Fig. 4 and Fig. 5.

TABLE 3. EVALUATION METRICS FOR MODELS OF PUMINER FRAMEWORK.

| Metric and Formula | Description |
|---|---|
| 1. $\text{Precision}_{PU-LB} = \frac{TP_P}{TP_P + TP_U + FP}$, $TP_P$ & $TP_U$: $|\hat{y}=1|$ in the positive and unlabelled sets, $\hat{y}$: predicted label, $y$: true label. | Minimum Precision when only $TP$ in the positive set is counted. $TP$ in $U$ is unknown, and thus not counted. |
| 2. $\text{Precision}_{PU-UB} = \frac{TP_P + \min(r \times |U|, TP_U + FP)}{TP_P + TP_U + FP}$, $r$: the ratio of security posts in $U$, $|U|$: size of $U$. | Maximum Precision when a model is assumed to obtain all security posts in $U$. |
| 3. $\text{Recall}_{PU} = \frac{TP_P}{TP_P + FN_P}$ ($\text{Recall}_{PU} \approx \text{Recall}_{PN}$) | Approximated Recall for PU learning [56] |
| 4. $\text{F1-Score}_{PU-LB} = \frac{2 \times \text{Precision}_{PU-LB} \times \text{Recall}_{PU}}{\text{Precision}_{PU-LB} + \text{Recall}_{PU}}$ | Minimum F1-Score for PU learning |
| 5. $\text{F1-Score}_{PU-UB} = \frac{2 \times \text{Precision}_{PU-UB} \times \text{Recall}_{PU}}{\text{Precision}_{PU-UB} + \text{Recall}_{PU}}$ | Maximum F1-Score for PU learning |
| 6. $\text{G-mean}_{PU} = \frac{|\hat{y}| \times R^2}{|\hat{y}=1|}$ | G-mean for PU Learning (Eq. (3)) |
| 7. $\text{Precision}_{PN} = \frac{TP}{TP + FP}$ | Precision on ground-truth posts |
| 8. $\text{Recall}_{PN} = \frac{TP}{TP + FN}$ | Recall on ground-truth posts |
| 9. $\text{F1-Score}_{PN} = \frac{2 \times \text{Precision}_{PN} \times \text{Recall}_{PN}}{\text{Precision}_{PN} + \text{Recall}_{PN}}$ | F1-Score on ground-truth posts |
| 10. $\text{G-mean}_{PN} = \sqrt{\text{Precision}_{PN} \times \text{Recall}_{PN}}$ | G-mean on ground-truth posts |
| 11. $\text{MCC}_{PN}$ $= \frac{TP \times TN - FP \times FN}{\sqrt{(TP+FP)(TP+FN)(TN+FP)(TN+FN)}}$ | Matthews correlation coefficient on ground-truth posts |

This confirmed our assumption in section III.D that the same-class posts would be close to each other in their respective regions, and thus making them distinguishable. Fig. 5 also demonstrated that there was only a tiny portion of security (purple) posts in the unlabelled set. This finding supported the argument that the unlabelled centroid could approximately represent the negative class in the centroid-based approach of our two-stage PU learning model.

Next, Table 4 summarizes the optimal validation results of each classifier. XGB with number of estimators = 500, max_depth = unlimited, and max_leaf_nodes = 300 performed the best (nearly 0.9) in every metric. No common optimal configuration was found for all models, suggesting that tuning process is important [15, 57, 58]. Also, the ratio of security to unlabelled posts was 1:1, and about 2.5% of the unlabelled set were security (see section IV.C), implying that 51.25% of the whole set was security. Thus, G-mean$_{PN}$ (XGB) = $\sqrt{\text{G-mean}_{PU} \times P(y=1)} = \sqrt{1.611 \times 0.5125} = 0.909$.

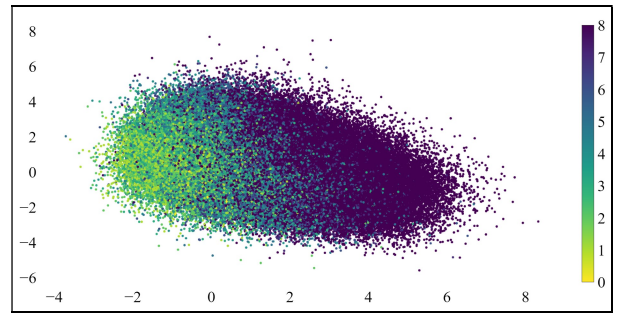

Fig. 4. Embeddings of security posts in our *development set*. **Note**: The color bar shows the number of keywords in our security list of a post.

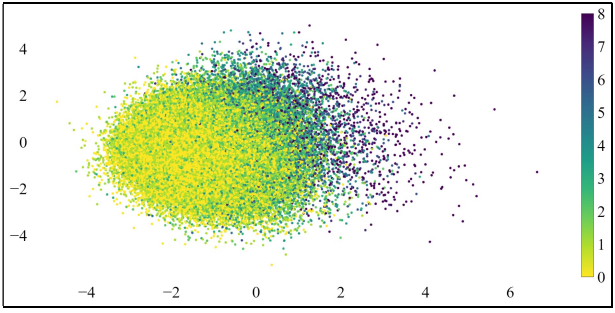

Fig. 5. Embeddings of unlabelled posts in our *development set*. **Note**: The color bar shows the number of keywords in our security list of a post.

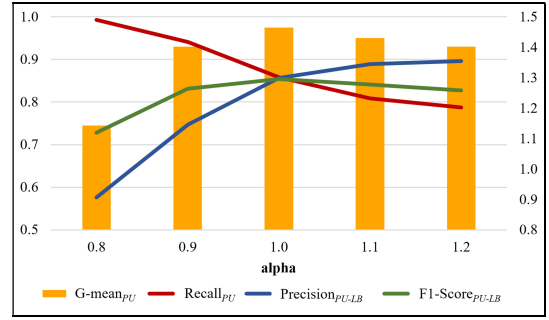

Fig. 6. Effect of alpha on the average performance of PU learning models. **Notes**: Right axis is for G-mean$_{PU}$. Left axis is for the other three metrics.

The worst approximated G-mean$_{PN}$ (KNN) was still 0.863, meaning that PUMiner performed well in overall (> 0.85 G-mean$_{PN}$) for retrieving security posts from Q&A sites.

TABLE 4. CROSS-VALIDATION PERFORMANCE OF THE OPTIMAL MODELS OF SIX CLASSIFIERS.

| Metric | LR | SVM | KNN | RF | XGB | LGBM |
|---|---|---|---|---|---|---|
| Recall$_{PU}$ | 0.869 | 0.869 | **0.893** | 0.862 | **0.893** | **0.893** |
| Precision$_{PU-LB}$ | 0.876 | 0.876 | 0.815 | 0.853 | **0.903** | 0.902 |
| Precision$_{PU-UB}$ | 0.901 | 0.901 | 0.838 | 0.878 | **0.928** | 0.927 |
| F1-Score$_{PU-LB}$ | 0.873 | 0.873 | 0.852 | 0.857 | **0.898** | 0.897 |
| F1-Score$_{PU-UB}$ | 0.885 | 0.885 | 0.864 | 0.870 | **0.910** | 0.910 |
| G-mean$_{PU}$ | 1.521 | 1.521 | 1.454 | 1.468 | **1.611** | 1.609 |

Next, Table 5 reports the performance of single (LR, SVM, and KNN) and ensemble (RF, XGB, and LGBM) models. Ensemble models were better than single counterparts in all metrics except Precision (mostly due to the poor performance of RF). Such differences were confirmed to be statistically significant by one-tailed non-parametric Mann-Whitney U-test [59] and non-negligible with Cohen's *d* effect size [60] (larger than 0.2 [61]). Precision$_{PU-UB}$ and F1-Score$_{PU-UB}$ had the same patterns as Precision$_{PU-LB}$ and F1-Score$_{PU-LB}$, respectively. We suggest that ensemble models (LGBM and XGB) should be preferred.

We also investigated the effect of alpha ($\alpha$) hyperparameter in the first stage of PUMiner on the performance (see Fig. 6). Higher values of alpha resulted in higher Precision and lower Recall. Intuitively, increasing alpha would move the selected negatives more towards the real positive class. It would then make the negative class noisier (i.e., more security posts in it), and produce more false negatives (low Recall) yet fewer false positives (high Precision). Alpha ≥ 1 also gave more stable (less changing) results than alpha < 1, and produced better average F1-Score and G-mean. This finding was confirmed when five out of six classifiers used alpha = 1.1 in their optimal models; whereas, KNN used alpha = 0.9. It might be because KNN is instance-based learning [21, 30], and thus it requires more refined/reliable negatives to learn the patterns. We recommend trying alpha ≥ 1 first for better baseline performance.

TABLE 5. P-VALUES OF HYPOTHESIS TESTS AND COHEN'S D EFFECT SIZES OF PERFORMANCE COMPARISON BETWEEN ENSEMBLE AND SINGLE MODELS.
**NOTES**: (*): H$_0$: ENSEMBLE ≤ SINGLE AND (**): H$_0$: SINGLE ≤ ENSEMBLE.

| Evaluation metric | Ensemble model | Single model | P-value | Cohen's *d* effect size |
|---|---|---|---|---|
| Recall$_{PU}$ | 0.914 | 0.838 | 8.07 x 10$^{-8}$ (*) | 0.785 |
| Precision$_{PU-LB}$ | 0.773 | 0.820 | 1.60 x 10$^{-4}$ (**) | 0.374 |
| F1-Score$_{PU-LB}$ | 0.827 | 0.810 | 9.12 x 10$^{-4}$ (*) | 0.287 |
| G-mean$_{PU}$ | 1.400 | 1.344 | 6.17 x 10$^{-4}$ (*) | 0.337 |

### B. RQ2: *How does PUMiner perform compared to other learning approaches for mining security posts from developer Q&A websites?*

**Motivation.** We can retrieve security posts using positive-only learning, a.k.a. novelty detection. One-class SVM (OC-SVM) [23, 62] is widely used for novelty detection by building a positive-class boundary with a max-margin SVM, while the region outside the boundary would be the negative class. Another novelty-detection method, Positive-Similarity Filtering (PSF) [24], classifies a new document as positive if the cosine similarity between its feature vector and that of any existing labelled positive document is higher than a threshold. RQ2 compares PUMiner with OC-SVM, PSF, and one-stage (centroid-based learning) PU (1-PU) models.

**Approach.** RQ2 compares the performance of the optimal PUMiner model in RQ1 on three unseen testing sets (Test$_{dev}$, Test$_{norm}$ and Test$_{hard}$ in section IV.A) against OC-SVM, PSF and 1-PU models. We optimize the baseline models on 90% data of the *development set* similarly to PUMiner. OC-SVM and PSF are only trained on positive samples and the same context-aware features are used for each model.

**Results.** Table 6 shows the optimal hyperparameters of the three baseline models. We compared such models with the optimal PUMiner model (XGB) reported in RQ1.

TABLE 6. HYPERPARAMETER TUNING FOR THE BASELINE MODELS.
**NOTE**: THE OPTIMAL CONFIGURATIONS ARE IN **BOLD**.

| Model | Hyperparameter | Values |
|---|---|---|
| 1-PU | alpha | 0.8, 0.9, **1.0**, 1.1, 1.2 |
| OC-SVM | kernel | **linear**, rbf |
|  | nu (Proportion of non-security posts) | **0.01**, 0.05, 0.1, 0.3, 0.5, 0.7, 0.9 |
| PSF | similarity threshold | **0.5**, 0.6, 0.7, 0.8, 0.9 |

On Test$_{dev}$, PUMiner performed the best in five out of six metrics (see Table 7), surpassing the second-best model (1-PU) by 16.7% in terms of G-mean$_{PU}$. OC-SVM had the highest Recall$_{PU}$, but it had very low Precision since it marked most of the posts as security.

TABLE 7. PERFORMANCE COMPARISON ON THE TEST$_{DEV}$ SET. **NOTE**: SCALED VALUES OF G-MEAN$_{PU}$ BY $P(Y=1) = 0.5125$ ARE IN THE PARENTHESES.

| Evaluation metric | PUMiner | 1-PU | OC-SVM | PSF |
|---|---|---|---|---|
| Recall$_{PU}$ | 0.892 | 0.783 | **0.985** | 0.650 |
| Precision$_{PU-LB}$ | **0.904** | 0.882 | 0.506 | 0.513 |
| Precision$_{PU-UB}$ | **0.929** | 0.910 | 0.518 | 0.532 |
| F1-Score$_{PU-LB}$ | **0.898** | 0.830 | 0.668 | 0.573 |
| F1-Score$_{PU-UB}$ | **0.910** | 0.842 | 0.679 | 0.585 |
| G-mean$_{PU}$ | **1.610** (**0.908**) | 1.380 (0.841) | 0.995 (0.714) | 0.665 (0.584) |

TABLE 8. PERFORMANCE COMPARISON ON THE TEST$_{NORM}$ AND TEST$_{HARD}$ SETS.
NOTE: THE RESULTS ON THE TEST$_{HARD}$ SET ARE IN PARENTHESES.

| Model | Recall$_{PN}$ | Precision$_{PN}$ | F1-Score$_{PN}$ | G-mean$_{PN}$ | MCC$_{PN}$ |
|---|---|---|---|---|---|
| PUMiner | **0.951** (0.745) | 0.943 (0.973) | **0.947** (**0.844**) | **0.947** (**0.852**) | **0.894** (**0.745**) |
| 1-PU | 0.849 (0.600) | **0.951** (**0.991**) | 0.897 (0.748) | 0.899 (0.771) | 0.810 (0.647) |
| OC-SVM | **0.951** (**0.948**) | 0.499 (0.489) | 0.654 (0.645) | 0.688 (0.681) | -0.012 (-0.13) |
| PSF | 0.377 (0.758) | 0.833 (0.692) | 0.519 (0.724) | 0.560 (0.724) | 0.360 (0.423) |

This behavior would not be useful in practice. The promising results of 1-PU model might be because of the reasonable separability of security and non-security posts illustrated in RQ1. PSF was the worst-performing model in this case.

Table 8 reports the performance on the labelled Test$_{norm}$ and Test$_{hard}$ sets, where the true estimation of traditional classification metrics can be computed. PUMiner obtained the best overall performance for both testing sets. On Test$_{norm}$, PUMiner outperformed the three baseline models in all metrics, except Recall$_{PN}$ for OC-SVM and Precision$_{PN}$ for 1-PU model. MCC$_{PN}$ of PUMiner was 0.084, 0.534 and 0.906 points higher than those of 1-PU, PSF and OC-SVM models, respectively. OC-SVM had the best Recall$_{PN}$, but its MCC$_{PN}$ was negative and even worse than a random classifier. 1-PU model had the best Precision$_{PN}$, yet its Recall$_{PN}$ was much lower than PUMiner. On Test$_{hard}$, PUMiner produced the best MCC$_{PN}$ value of 0.745, which was 0.098, 0.322 and 0.875 points superior to those of 1-PU, PSF and OC-SVM models, respectively. We also recorded that OC-SVM had the highest Recall$_{PN}$ yet poor Precision$_{PN}$, and vice versa for 1-PU model on Test$_{hard}$. Also, PSF was the worst model with the lowest F1-Score and G-mean for identifying security posts. In addition, the results were reasonable that Test$_{norm}$ > Test$_{dev}$ and Test$_{hard}$ < Test$_{dev}$ since Test$_{dev}$ contains both normal and borderline test cases, proving that our sample selection was not biased. These findings have reinforced the potential value of using unlabelled data and the second stage of PUMiner. Importantly, the heuristics in section III.C could not correctly detect the type of any post in Test$_{hard}$, which highlights the usefulness of PUMiner for reliable and large-scale security posts mining from Q&A websites.

### C. RQ3: *How do the source and size of training data affect the performance of PUMiner?*

**Motivation.** The major advantage of PU learning is working with partially labelled positive samples and a large amount of unlabelled data, which cannot be handled by traditionally fully-supervised learning. RQ3 investigates the impact of data source and size on the performance of PUMiner.

**Approach.** We first investigate different security datasets: set$_{tag}$, set$_{content}$ and Security StackExchange (see section IV.A). We then change the amount of labelled positive and unlabelled data to evaluate our PUMiner in different scenarios. We also compare the optimal PUMiner model with a fully supervised model that considers all unlabelled data as pseudo-negatives (PPN) to investigate the impact of the unlabelled set on the model performance. We use the same optimal model configurations of PUMiner and features for PPN. We also only report the results on the Test$_{norm}$ and Test$_{hard}$ sets as Test$_{dev}$ changes with each value of the positive ratio, which makes the results incomparable.

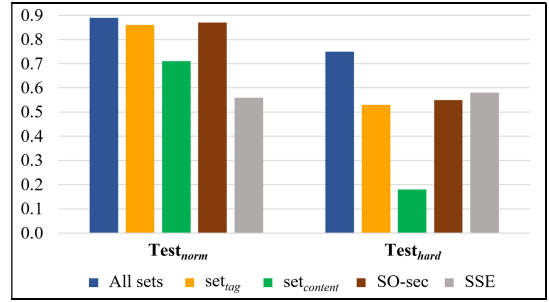

Fig. 7. MCC$_{PN}$ of different positive datasets on the Test$_{norm}$ and Test$_{hard}$ sets.
Notes: SO-sec is our security dataset extracted from Stack Overflow (set$_{tag}$ + set$_{content}$). SSE is Security StackExchange only.

**Results.** Fig. 7 shows that Stack Overflow (SO) was more helpful for identifying the security posts in Test$_{norm}$ set, while Security StackExchange (SSE) contributed more information to predicting the posts in Test$_{hard}$ set. The security posts extracted from SO (SO-sec) using our heuristics were more refined, while discussion on SSE was less restricted due to the informal nature of Q&A forums. Thus, the normal security posts were better captured using the patterns of SO-sec. In contrast, the security context of the posts in Test$_{hard}$ was less explicit; hence, the more diverse content of SSE could better cover such borderline cases. Overall, combining both SO-sec and SSE sources has been useful for retrieving security posts in both normal and extreme scenarios. Set$_{tag}$ also outperformed set$_{content}$ in both cases, probably due to the nearly four-time difference in size (102,046 vs. 27,424).

Next, when varying the ratio of security to unlabelled posts, PUMiner consistently had higher MCC$_{PN}$ than PPN (see Fig. 8). This has demonstrated that fully-supervised learning is not as effective as PUMiner when there is unlabelled data, confirming our statement made in section III.D. PPN was also seen to approach PUMiner in Test$_{norm}$ as the ratio increased. A reason might be that the result became more dominated by the increasing size of the positive set than the same positive posts in the unlabelled set. It should be noted that PUMiner also outperformed 1-PU, OC-SVM and PSF models for each value of the positive ratio.

We also examined the case when the unlabelled set increased by ten and 100 times (i.e., 1,346,926 and 12,999,226 posts). For the 1:10 case, PUMiner achieved MCC$_{PN}$ of 0.78 and 0.46 for Test$_{norm}$ and Test$_{hard}$ sets, respectively. With 100-time noisier data, MCC$_{PN}$ values of PUMiner were still 0.65 (Test$_{norm}$) and 0.35 (Test$_{hard}$), significantly surpassing PPN by 160% and 106%, respectively. This result demonstrates the robustness of PUMiner for information retrieval on Q&A sites even with a large amount of unlabelled data.

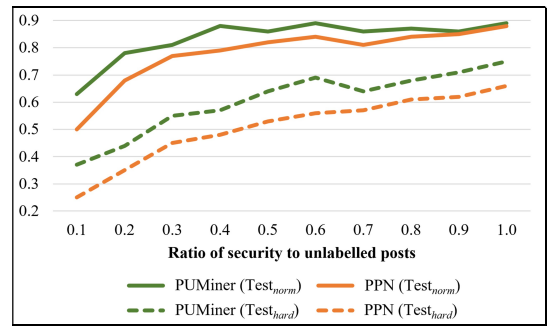

Fig. 8. MCC$_{PN}$ of PUMiner and PPN on the Test$_{norm}$ and Test$_{hard}$ sets with different ratios of security to unlabelled posts in the *development set*.

## VI. DISCUSSION

### A. PUMiner framework and beyond

We have demonstrated that PUMiner is the best security-retrieval model in section V. Moreover, PUMiner is efficient as it took only 1,400 seconds to learn 360,000 posts in the *development set*, and less than four seconds to predict the security relevance for 770 posts in each of the sets (Test$_{norm}$ and Test$_{hard}$). PUMiner also allows incremental training on new posts, making it suitable to keep up with the fast and large-scale update of Q&A websites. Using the optimal PUMiner model in RQ1 and our heuristics with $kw\_count \geq 5$ and $kw\_ratio \geq 0.03$, we have identified and released 104,024 new security posts on Stack Overflow (only 1% *FP* in 385 random posts) at [25] for future software security study. PUMiner can be extended to 170+ other topics on StackExchange [63], many of which are related to Software Engineering such as mobile, game, and web development.

We still found several misclassification patterns that can improve PUMiner in future work. For *FP*s, PUMiner struggled with the non-security posts about tools appearing in the security context, e.g., SO-1967980/44775539 about *Spring (boot)* framework were confused with *Spring security*. For *FN*s, PUMiner tended to miss security posts with implicit security context (e.g., SSE-34394 about *REcon* security conference or SO-37183524 about *OAuth* yet embedded in links and function calls), or about infrequent (< 1%) security tools or techniques (e.g., SSE-134794 about *Snort* or SO-16455365 about *Diffie-Hellman*). *Snort* and *Diffie-Hellman* only appeared in 432 and 1,426 posts, respectively, out of 181,696 total posts in our security dataset in section IV.A.

### B. Threats to validity

One potential threat to validity is with our data selection. We did use random sampling to avoid any bias. Our testing sets still involved human judgment, but we performed cross-checking with three people to make sure that the selected ones were as accurate and objective as possible.

Another threat is that our optimal hyperparameters of PUMiner might not produce the absolute highest results. However, it is nearly impossible to try all the combinations. To minimize this threat, we tried to follow the practices of previous work, and then ran our models multiple times with stratified 10-fold cross-validation. We also showed that PUMiner performed much better than other existing models.

Our generalizability is also a concern. We used the largest training dataset, statistically significant size of testing sets [29], and confirmed our results with a statistical confidence level > 95%. We also published our code and models at [25] to support future reuse and extension of PUMiner.

## VII. RELATED WORK

### A. Stack Overflow for Software Engineering research

Stack Overflow (SO) has long been used to support various Software Engineering (SE) tasks such as API usage/documentation [64-66], code comment generation [67], code debugging/fixing [68, 69] and code clone detection [70, 71]. Among the studies, [64] was evaluated in the context of unlabelled data; however, their labelled dataset contained both the positive and negative classes. Thus, it was less challenging than our study since we only have labelled positive posts and no negative class. Other studies identified topics of different domains [26, 72-74] on SO using unsupervised models (e.g., Latent Dirichlet Allocation (LDA) [75]), requiring no label. However, it is nearly impossible to directly guide such topic models to retrieve domain-specific posts. In addition, tag recommendation on Q&A websites is related to our task, but most of the existing techniques [76-78] required fully labelled data (post-tag pairs). It is also very difficult to obtain all relevant tags of a domain.

### B. Mining security knowledge from open sources

Yang et al. [12] identified and analyzed thirty security topics on SO using LDA [75]. However, they only obtained 30,054 security posts, which were at least six times fewer than our work. Such missing posts may lead to an overlook of current security issues and countermeasures. There is also active research on software vulnerabilities analytics using CVE/NVD [15, 79, 80], but these sources do not support online discussion for developers, and security issues and countermeasures may not be updated in time [81, 82]. Pletea et al. [14] proposed a list of security keywords to perform sentimental analysis of security discussions on GitHub. However, we have shown that this list still misses many security keywords (e.g., *buffer-overflow*, *directory-traversal*, and *vulnerability*). Our work has proposed an updated list (see Appendix) that has two times more security keywords than [14]. Recently, [83] and [24] have used novelty detection methods to retrieve security-related content. We have shown that PUMiner utilizing unlabelled data performed much better than one-class SVM and positive-similarity filtering used in these two studies.

## VIII. CONCLUSIONS AND FUTURE WORK

We have proposed a novel framework, PUMiner, to automatically mine security content from Q&A websites where only some positive posts are known and the rest is unlabelled. Extensive experiments on more than 17.3 million posts from Stack Overflow and Security StackExchange have shown that PUMiner is a promising approach with at least 0.85 validation F1-Score and G-mean. On unseen testing posts, PUMiner significantly outperformed one-class SVM, positive-similarity filtering and one-stage PU models, respectively. PUMiner also predicted the cases where keyword matching totally missed with an MCC of 0.745. Notably, with only 1% labelled positive posts, PUMiner still achieved an MCC of 0.65, which was 160% better than fully-supervised learning. Overall, PUMiner helps to reduce the human effort to retrieve information from large-scale and mostly unlabelled open sources in the SE community.

In the future, we would like to extend our PUMiner to other related domains such as security vulnerability analytics. We also plan to develop more sophisticated models and incorporate them into PUMiner to improve the effectiveness of information retrieval on developer Q&A websites.

ACKNOWLEDGEMENTS

The work has been supported by the Cyber Security Research Centre Limited whose activities are partially funded by the Australian Government's Cooperative Research Centres Programme. This work has also been supported with supercomputing resources provided by the Phoenix HPC service at the University of Adelaide.

## APPENDIX

234 security keywords used in this work: access control, access role, adware, adversarial, malware, spyware, ransomware, aes, antivirus, asset, audit, authority, authorise, availability, bitlocker, biometric, blacklist, botnet, buffer overflow, burp, ctf, capture the flag, cbc, certificate, checksum, cipher, clearance, confidential, clickfraud, clickjacking, cloudflare, cookie, crc, credential, crypt, csrf, ddos, danger, data exfiltrate, data exfiltration, data breach, decode, defence, defense, defensive programming, delegation, denial of service, diffie hellman, directory traversal, disclose, disclosure, dmz, dotfuscator, dsa, ecdsa, encode, encrypt, escrow, exploit, evil twin, fingerprint, firewall, forge, forgery, fuzz, fraud, gnupg, gss api, hack, hash, hijacking, hmac, honeypot, honey pot, hsm, inject, insecure, integrity, intrusion, intruder, ipsec, kerberos, ldap, login, metasploit, meterpreter, malicious, md5, nessus, nonce, nss, oauth, obfuscate, openssl, openssh, openvas, open auth, open redirect, openid, owasp, password, pbkdf2, pci dss, pgp, phishing, pki, privacy, private key, privilege, privilege escalation, permission escalation, public key, pcidss, pen test, penetration test, protect, rainbow table, rbac, rc4, repudiation, rfc 2898, rijndael, rootkit, rsa, safe, salt, saml, sanitise, sandbox, scam, script kiddie, scripting, security, sftp, sha, shell code, shib boleth, signature, signed, signing, single sign on, smart assembly, smartassembly, snif, spam, spnego, spoof, ssh, ssl, sso, steganography, tampering, theft, threat, tls, transport, tunneling, tunnelling, trojan, trust, two factor, user account, user name, violate, validate, virus, white list, worm, x 509, x.509, xss, xxe, ssrf, zero day, 0 day, zombie computer, attack, vulnerability, attack vector, authentication, cross site, sensitive information, leak, information exposure, path traversal, use after free, double free, man in the middle, man in middle, mitm, poisoning, unauthorise, dot dot slash, bypass, session fixation, forced browsing, nvd, cwe, cve, capec, cpe, common weakness enumeration, common platform enumeration, crack, xml entity expansion, http parameter pollution, eavesdropping, cryptanalysis, http flood, xml flood, udp flood, tcp flood, tcp syn flood, steal, ssl flood, j2ee misconfiguration, asp.net misconfiguration, improper neutralisation, race condition, null pointer dereference, untrusted pointer dereference, trap door, back door, time bomb, xml bomb, logic bomb, captcha, deadlock, missing synchronisation, incorrect synchronisation, improper synchronisation, illegitimate, breach, sql injection